\documentclass{ws-procs9x6}

\begin{document}

\title{
Numerical study of the equation of state for two flavor QCD at finite density
\footnote{\uppercase{P}resented by \uppercase{S. E}jiri.
\uppercase{T}his work is supported by 
\uppercase{BMBF} under grant \uppercase{N}o.06\uppercase{BI}102, 
\uppercase{DFG} under grant \uppercase{FOR} 339/2-1 
and \uppercase{PPARC} grant \uppercase{PPA}/a/s/1999/00026.}}

\author{S.~Ejiri\rlap,$^{\lowercase{\rm a}}$ 
C.R.~Allton\rlap,$^{\lowercase{\rm b}}$ 
S.J.~Hands\rlap,$^{\lowercase{\rm b}}$ 
O.~Kaczmarek\rlap,$^{\lowercase{\rm a}}$ 
F.~Karsch\rlap,$^{\lowercase{\rm a}}$ 
E.~Laermann\rlap,$^{\lowercase{\rm a}}$ \lowercase{and} 
C.~Schmidt$^{\lowercase{\rm a}}$}

\address{$^{\lowercase{\rm a}}$Fakult\"{a}t f\"{u}r Physik, 
Universit\"{a}t Bielefeld, D-33615 Bielefeld, Germany \\
$^{\lowercase{\rm b}}$Department of Physics, University of 
Wales Swansea, Singleton Park, Swansea, SA2 8PP, U.K.
}


\maketitle

\abstracts{
We discuss the equation of state for 2 flavor QCD at 
non-zero temperature and density. 
Derivatives of $\ln Z$ with respect to 
quark chemical potential $\mu_q$ up to fourth order are calculated, 
enabling estimates of the pressure, quark number density and 
associated susceptibilities as functions of $\mu_q$ via a Taylor 
series expansion. 
It is found that the fluctuations in the quark number density 
increase in the vicinity of the phase transition temperature and 
the susceptibilities start to develop a pronounced peak 
as $\mu_q$ is increased. 
This suggests the presence of a critical endpoint in the 
$(T, \mu_q)$ plane.}

\section{Introduction}
\label{sec:intro}

Remarkable progress for QCD thermodynamics has been recently made 
by numerical studies of lattice QCD with small but non-zero baryon 
density\cite{MNN}. The pseudocritical line has been investigated 
in the low density regime\cite{Fod,us02,dFP}. 
The phase transition for 2 flavor QCD is known to be crossover 
at $\mu_q=0$ and expected to become a first order phase 
transition at a critical endpoint, and it might be 
possible to detect the endpoint experimentally via event-by-event 
fluctuations in heavy-ion collision experiments.
 
In this report, we discuss the equation of state 
at non-zero baryon number density. 
The study of the equation of state gives the most basic information 
for the experiments. Quantitative calculations of thermodynamic quantities 
such as pressure and energy density are indispensable. 
In particular, since the number density fluctuation should be large 
around the endpoint\cite{RSS}, 
the susceptibility of the quark number density, 
given by the second derivative of pressure with respect to $\mu_q$, 
is an important quantity\cite{AHMJK}. 

Several studies of the quark number susceptibility 
have been performed at $\mu_q=0$ \cite{Gott}. 
Moreover, measurements of pressure and energy density at 
$\mu_q \neq 0$ were done by Ref.~\refcite{FKS} using the reweighting method, 
which allow the investigation of thermodynamic properties at 
non-zero baryon density. This approach, however, does not work 
for large $\mu_q$ and large lattice size due to the sign problem. 

Here, we adopt the following strategy. 
We compute the derivatives of physical quantities with respect to $\mu_q$ 
at $\mu_q=0$, and determine the Taylor expansion coefficients in 
$\mu_q$\cite{us02,TARO}, in which the sign problem does not arise. 
Because the pressure is an even function of $\mu_q$, 
the $\mu_q^2$-term is leading and $\mu_q^4$-term is the next to leading,
and in fact only these two terms are non-zero 
in the high temperature limit.
We compute the Taylor expansion coefficients up to fourth order. 
Using the second derivative of pressure (energy density), 
we investigate the relation between the line of constant pressure 
(energy density) and the phase transition line\cite{us02}. 
The fourth order term enables us to evaluate the $\mu_q$-dependence 
of the quark number susceptibility near $\mu_q=0$ . 
By estimating the change of the susceptibility, 
we discuss the possibility of the existence of the critical 
end point in the phase diagram of $T$ and $\mu_q$\cite{us03}.

\section{Taylor expansion in $\mu_q$}
\label{sec:expansion}

Pressure is given in terms of the grand partition function 
$Z(T, V, \mu_q)$ by
$p/T^4=(1/VT^3)\ln Z.$
However, the direct calculation of $\ln Z$ is difficult, hence 
most of the work done at $\mu_q=0$ for the calculation of pressure 
is done by using the integral method\cite{KLP,CPPACS}, 
where the first derivative of 
pressure is computed by simulations, and the pressure is obtained by 
integration along a suitable integral path. 
For $\mu_q \neq 0$, direct Monte Carlo simulation is not applicable; 
in this case we proceed by computing higher order derivatives of 
pressure with respect to $\mu_q/T$ at $\mu_q=0$, and then 
estimate $p(\mu_q)$ using a Taylor expansion, 
\begin{eqnarray}
\left. \frac{p}{T^4} \right|_{T,\mu_q} 
= \left. \frac{p}{T^4} \right|_{T,0} 
+ \sum_{n=1}^{\infty} c_n(T) \left( \frac{\mu_q}{T} \right)^n ,
\label{eq:taylorcont}
\end{eqnarray}
where $c_n=(1/n!) \partial^n(p/T^4)/\partial(\mu_q/T)^n |_{\mu_q=0}$.
These derivatives can be computed by the random noise method, which 
saves CPU time, and also it can be proved that 
the odd terms are exactly zero. 
Furthermore, we do not need simulations at $(T, \mu_q)=(0,0)$ 
for the subtraction to normalize the value of $p$, since 
the derivatives of $p|_{T=0,\mu_q=0}$ with respect to $\mu_q$ are, 
of course, zero. This also reduces CPU time.

\begin{figure}[t]
\centerline{\epsfysize=2.1in\epsfbox{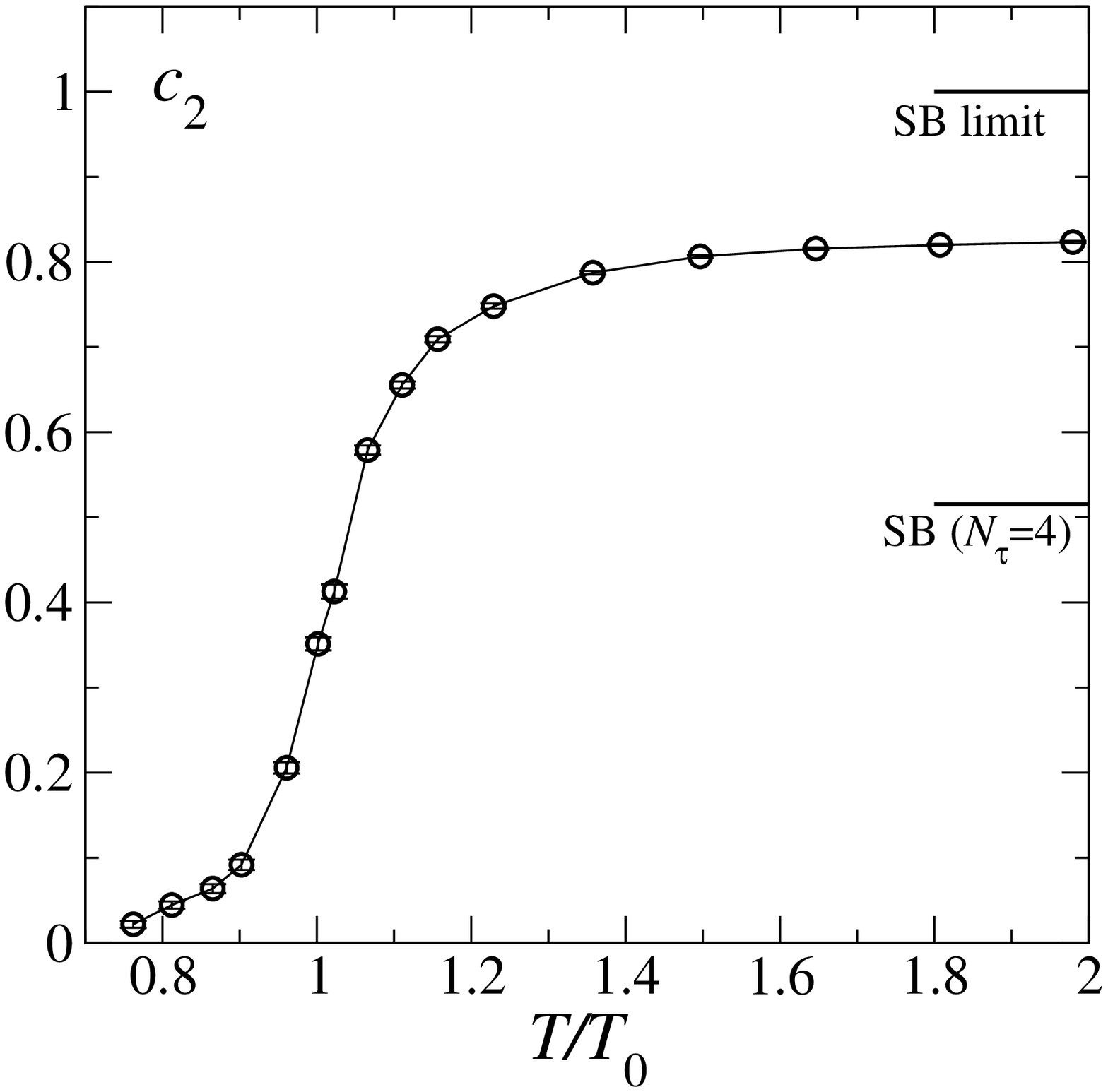}
            \epsfysize=2.1in\epsfbox{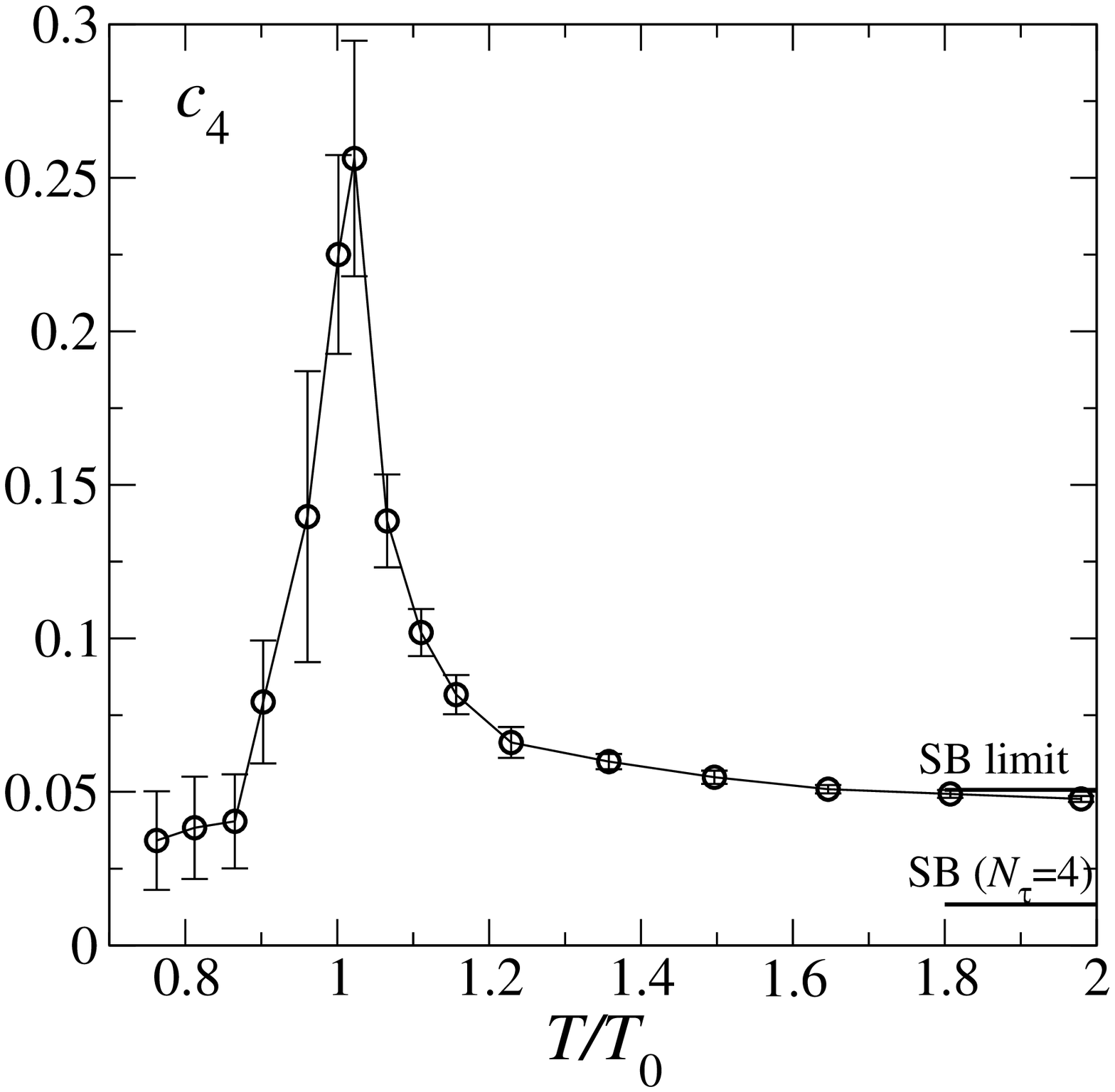}}   
\caption{
Coefficients of Taylor expansion, $c_2$ (left) and $c_4$ (right).
$T_0$ is $T_c$ at $\mu_q=0$.
\label{fig:c2c4}}
\end{figure}

The quark number density $n_q$ and quark number susceptibility 
$\chi_q$ are obtained by the derivatives of pressure; 
$n_q/T^3=\partial(p/T^4)/\partial(\mu_q/T)$,  
$\chi_q/T^2 = \partial^2 (p/T^4)/\partial (\mu_q/T)^2$. 
We compute the pressure up to $O(\mu_q^4)$ using 2 flavors of 
p4-improved staggered fermion\cite{HKS} at $ma=0.1$ 
on a $16^3 \times 4$ lattice. 
Then the quark number density and quark number susceptibility are 
obtained up to $O(\mu_q^3)$ and $O(\mu_q^2)$, respectively. 
The details of the simulations are given in Ref.~\refcite{us03}. 
In Fig.~\ref{fig:c2c4}, we plot the data for $c_2$ (left) and 
$c_4$ (right). Both of them are very small at low temperature and 
approach the Stefan-Boltzmann (SB) limit in the high temperature 
limit, as we expected. 
The remarkable point is a strong peak of $c_4$ around $T_c$. 

We can understand this peak via two arguments. 
One is a prediction from the hadron resonance gas model\cite{KRT}, 
which is an effective model of the free hadron gas in the low temperature 
phase. 
The model study predicts $c_4/c_2=0.75$ and our results are consistent 
with this prediction for $T < T_c$; in fact, as $T$ increases $c_4/c_2$ 
remains constant until $T \approx T_c$, 
whereupon it approaches the SB limit. 

The other point is from a discussion of the convergence radius of 
the Taylor expansion. 
We expect that the crossover transition at $\mu_q=0$ changes to 
first order transition at a point $\mu_q/T_c \sim O(1)$\cite{Fod,FK03}. 
Then, the analysis by Taylor expansion 
must break down in that regime, i.e. the convergence radius should be 
smaller than the value of $\mu_q/T$ at the critical endpoint.
We define estimates for the convergence radius by 
$\rho_n \equiv \sqrt{|c_n / c_{n+2}|}$.
We compute $\rho_0$ and $\rho_2$ from $c_0 \equiv p/T^4(0)$, $c_2$ and $c_4$. 
It is found that both $\rho_0$ and $\rho_2$ are quite large at 
high temperature as expected from the SB limit,
$\rho_2^{SB} \simeq 2.01, \rho_4^{SB} \simeq 4.44.$ On the other hand, 
around $T_c$, these are $O(1)$, since $c_2$ and $c_4$ are of the same order, 
so that our results around $T_c$ suggest a singular point in the 
neighborhood of $\mu_q/T_c =1$.

\section{Equation of state at $\mu_q \neq 0$} 
\label{sec:eos}
Next, we calculate pressure and quark number susceptibility in a range 
of $0 \leq \mu_q/T \leq 1$ which is within the radius of convergence 
discussed above, using the data of $c_2$ and $c_4$; 
$\Delta(p/T^4) \equiv p(T,\mu_q)/T^4-p(T,0)/T^4=
c_2(\mu_q/T)^2+c_4(\mu_q/T)^4+O(\mu_q^6)$, 
and $\chi_q/T^2=2c_2+12c_4(\mu_q/T)^2+O(\mu_q^4)$. 

\begin{figure}[t]
\centerline{\epsfysize=2.0in\epsfbox{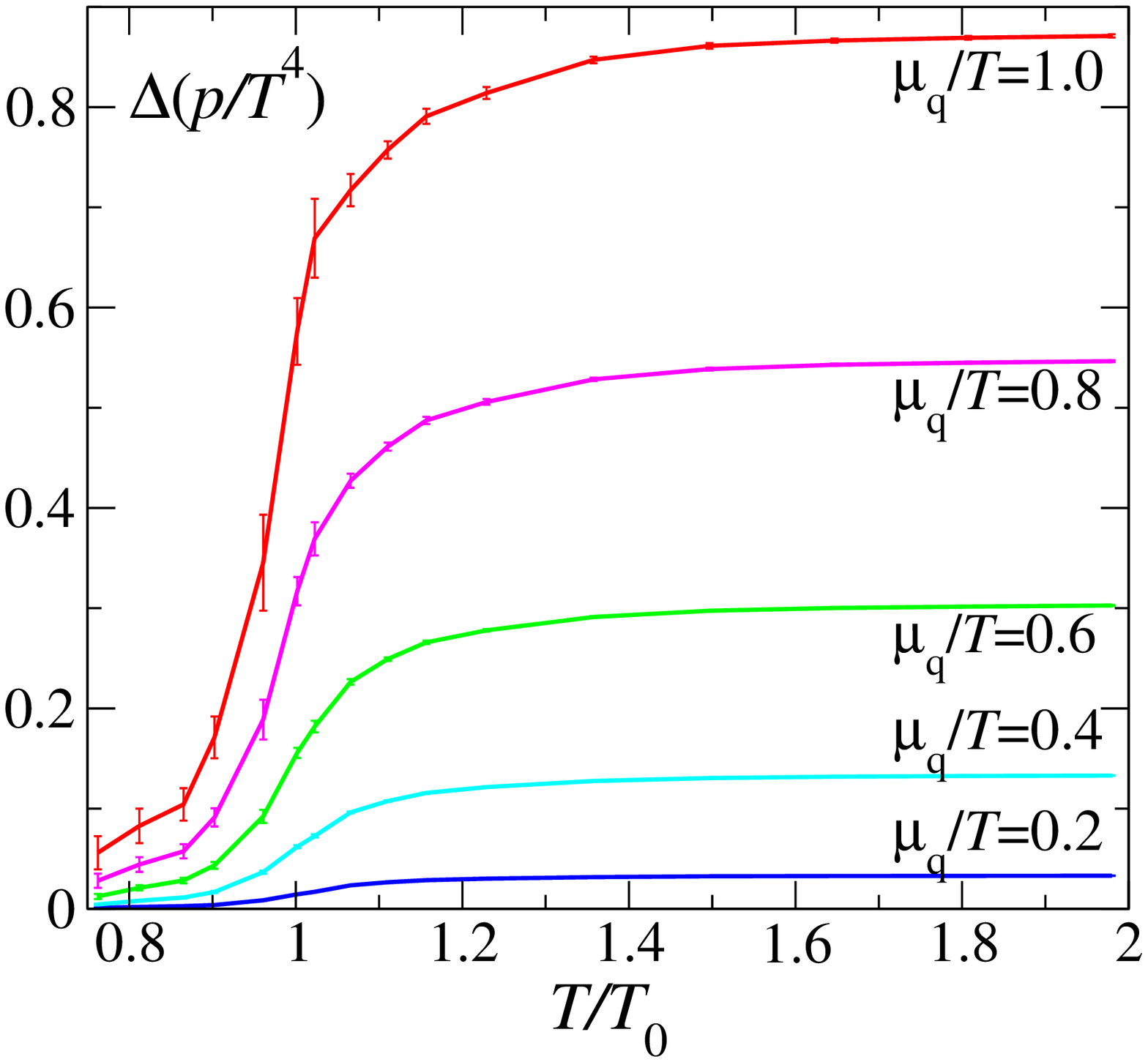}
            \epsfysize=2.0in\epsfbox{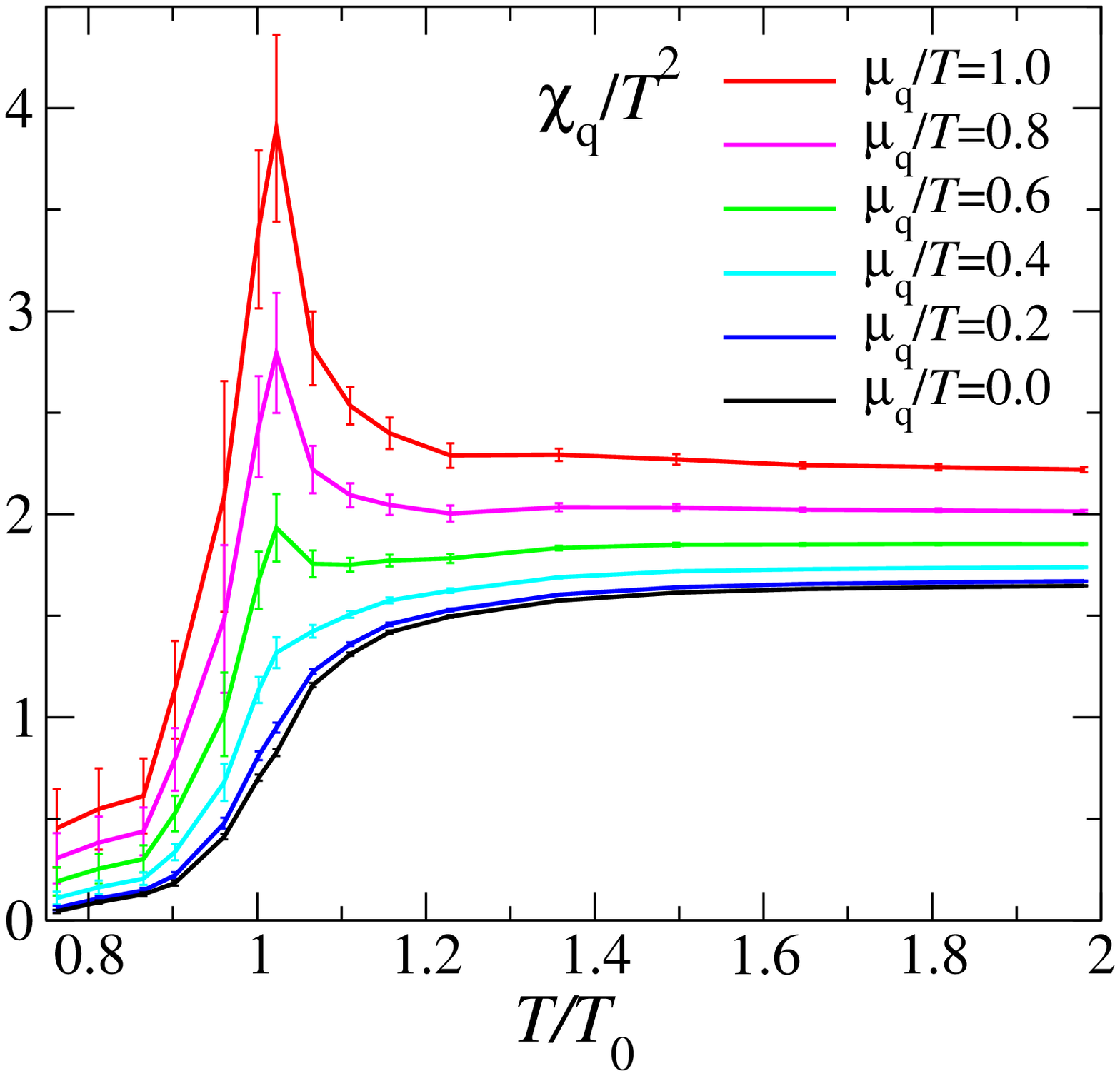}}   
\caption{Difference of pressure from $\mu_q=0$ (left) and
Quark number susceptibility (right) as a function of $T$ 
for each fixed $\mu_q/T$.
$T_0$ is $T_c$ at $\mu_q=0$.
\label{fig:pres}}
\end{figure}

We draw $\Delta (p/T^4)$ for each fixed $\mu_q/T$ in Fig.~\ref{fig:pres} 
(left) and find that the difference from $p|_{\mu_q=0}$ is very small 
in the interesting regime for heavy-ion collisions, 
$\mu_q/T \approx 0.1$ (RHIC) and $\mu_q/T \approx 0.5$ (SPS), 
in comparison with the value at $\mu_q=0$, e.g. the SB value for 2 flavor QCD 
at $\mu_q=0$: $p^{SB}/T^4 \simeq 4.06$. The effect of non-zero quark density 
on pressure at $\mu_q/T=0.1$ is only $1\%$. 
Also, the result is qualitatively consistent with that 
of Ref.~\refcite{FKS} obtained by the reweighting method. 

Moreover, together with the data of derivative with respect to $\beta$, 
we can discuss the lines of constant pressure [energy density] at $T_c$. 
From ${\rm d} p(T,\mu^2)=0$, the slope at $\mu_q=0$ is 
\begin{eqnarray}
\frac{{\rm d} T}{{\rm d} (\mu_q^2)} = \left. 
- \frac{\partial(p/T^4)}{\partial (\mu_q^2)} 
\right/ \left( \frac{\partial (p/T^4)}{\partial T} + \frac{4p}{T^5} \right).
\end{eqnarray}
We obtain
$T^2 \partial^2(p/T^4)/\partial \mu_q^2 = 0.693(5)$, and 
use the data in Ref.~\refcite{KLP}, 
$p/T^4 = 0.27(5)$, 
$T \partial (p/T^4)/\partial T =
-[a^{-1}(\partial a/\partial \beta)]^{-1} 
(\partial(p/T^4)/\partial \beta) = 2.2(6)$, 
at $\beta_c$ for $ma=0.1$. 
The same calculation is also performed for the energy density.
We find that the slope of the constant pressure [energy density] line is 
$T({\rm d} T / {\rm d} (\mu_q^2)) = -0.107(22) [-0.087(23)].$ 
Since the slope of $T_c$ in terms of $\mu_q^2$ is 
$T_c ({\rm d} T_c / {\rm d} (\mu_q^2)) 
= -0.07(3)$ \cite{us02}, 
this result suggests that the line of constant pressure or 
energy density is parallel with the phase transition line, and 
the $\mu_q$-dependence of the pressure or energy density at 
$T_c (\mu_q)$ is very small\cite{us02}. 

Figure \ref{fig:pres} (right) is the result for $\chi_q/T^2$ at fixed 
$\mu_q/T$. 
We find a pronounced peak for $m_q/T > 0.5$, whereas $\chi_q$ does 
not have a peak for $\mu_q=0$.
This suggests the presence of a critical endpoint in the $(T,\mu_q)$ plane.

This discussion can be easily extended to the charge fluctuation 
$\chi_C$.
The spike of $\chi_C$ at $T_c$ is weaker than that of 
$\chi_q$, which means the increase of the charge fluctuation is smaller 
than that of the number fluctuation as $\mu_q$ increases\cite{us03}.

\end{document}